\documentclass[11pt]{article}
\usepackage{axodraw}
\usepackage{epsfig}
\usepackage{amsfonts}
\usepackage{amsmath}
\usepackage{bbm}
\allowdisplaybreaks[1]

\input pix.sty

\newcommand{\gB}{g_\rmii{B}}
\newcommand{\gammaE}{\gamma_\rmii{E}}
\renewcommand{\eq}{eq.~}
\renewcommand{\eqs}{eqs.~}
\renewcommand{\se}{sec.~}

\renewcommand{\fig}{fig.~}
\renewcommand{\figs}{figs.~}

\newcommand{\tinymsbar}{{\overline{\mbox{\tiny\rm{MS}}}}}
\newcommand{\Lambdamsbar}{{\Lambda_\tinymsbar}}

\newcommand{\Nf}{N_{\rm f}}
\newcommand{\Nc}{N_{\rm c}}

\newcommand{\Tc}{T_{\rm c}}

\newcommand{\rmO}{{\mathcal{O}}}
\newcommand{\bmu}{\bar\mu}
\newcommand{\CA}{\Nc}
\newcommand{\CF}{C_\rmii{F}}
\newcommand{\TF}{T_\rmii{F}}

\def\lsi{\raise0.3ex\hbox{$<$\kern-0.75em\raise-1.1ex\hbox{$\sim$}}}
\def\gsi{\raise0.3ex\hbox{$>$\kern-0.75em\raise-1.1ex\hbox{$\sim$}}}
\newcommand{\lsim}{\mathop{\lsi}}
\newcommand{\gsim}{\mathop{\gsi}}

\newcommand{\nF}{n_\rmii{F}}
\newcommand{\nB}{n_\rmii{B}}
 \renewcommand{\nF}[1]{n_\rmii{F{#1}}}
 \renewcommand{\nB}[1]{n_\rmii{B{#1}}}
\newcommand{\rmii}[1]{{\mbox{\tiny\rm{#1}}}}
\newcommand{\rmiii}[1]{{\mbox{\tiny{$\scriptstyle{\rm#1}$}}}}

\newcommand{\Tint}[1]{{\hbox{$\sum$}\!\!\!\!\!\!\!\int\,}_{\!\!\!\!\raise-0.9ex\hbox{$\scriptstyle{#1}$}}}
\newcommand{\Tinti}[1]{{{\Sigma}\!\!\!\!\raise0.3ex\hbox{$\int$}_\rmii{${#1}$}}}

\newcommand{\bi}{\begin{itemize}}
\newcommand{\ei}{\end{itemize}}

\newcommand{\hide}[1]{ }
\newcommand{\bsl}[1]{\,\slash\!\!\!\!{#1}\,}

\newcommand{\ff}{\rmi{\sl f\,}}
\def\TAsc(#1,#2)(#3,#4,#5)%
{\SetWidth{2.0}\CArc(#1,#2)(#3,#4,#5)\SetWidth{1.0}}
\def\Lwidth{3}

\def\TAgl(#1,#2)(#3,#4,#5){\SetWidth{2.0}\PhotonArc(#1,#2)(#3,#4,#5){\Lwidth}%
{6.283 #3 mul 360 div #4 #5 sub #4 #5 sub mul sqrt mul Tdensity mul}%
\SetWidth{1.0}}
\def\TLgl(#1,#2)(#3,#4){\SetWidth{2.0}\Photon(#1,#2)(#3,#4){\Lwidth}
{#1 #3 sub #1 #3 sub mul #2 #4 sub #2 #4 sub mul add sqrt Tdensity mul}%
\SetWidth{1.0}}
\newcommand{\piC}[1]{\;\parbox[c]{40pt}{\begin{picture}(120,60)(0,-20)
\SetWidth{1.0}\SetScale{0.35} #1 \end{picture}}\;}
\def\ConnectedA(#1,#2,#3){\piC{#1(60,-15)(75,34,146) #2(60,75)(75,214,326)%
 #3(60,60)(20,190,350)%
 \GBoxc(0,30)(10,10){1} \GBoxc(120,30)(10,10){1}%
  }}
\def\ConnectedB(#1,#2,#3){\piC{#1(60,-15)(75,34,146) #2(60,75)(75,214,326)%
 #3(60,60)(60,0)%
 \GBoxc(0,30)(10,10){1} \GBoxc(120,30)(10,10){1}%
  }}
\def\ConnectedC(#1,#2){\piC{#1(60,-15)(75,34,146) #2(60,75)(75,214,326)%
 \GBoxc(0,30)(10,10){1} \GBoxc(120,30)(10,10){1}%
  }}
\def\ConnectedD(#1,#2){\piC{#1(60,-15)(75,34,146) #2(60,75)(75,214,326)%
 \GBoxc(0,30)(10,10){1} \GBoxc(120,30)(10,10){1}%
 \SetWidth{2.0} 
 \Line(55,55)(65,65)%
 \Line(55,65)(65,55)
  }}

\makeatletter \@addtoreset{equation}{section} \makeatother
\renewcommand{\theequation}{\arabic{section}.\arabic{equation}}
\makeatletter
\renewcommand\section{\@startsection {section}{1}{\z@}%
                                   {-5.5ex \@plus -1ex \@minus -.2ex}
                                   {2.3ex \@plus.2ex}%
                                   {\normalfont\large\bfseries}}
\renewcommand\subsection{\@startsection{subsection}{2}{\z@}%
                                     {-3.25ex\@plus -1ex \@minus -.2ex}%
                                     {1.5ex \@plus .2ex}%
                                     {\normalfont\normalsize\bfseries}}
\renewcommand\thesection {\@arabic\c@section}
\renewcommand\thesubsection   {\thesection.\@arabic\c@subsection}
\renewcommand{\@seccntformat}[1]{%
\csname the#1\endcsname.\hspace{1.0em}}
\makeatother

\begin{document}

\flushbottom

\begin{titlepage}

\begin{flushright}
\vspace*{1cm}
\end{flushright}
\begin{centering}
\vfill

{\Large{\bf
 Charm mass effects in bulk channel correlations
}} 

\vspace{0.8cm}

Y.~Burnier$^{\rm a}$ 
 and 
M.~Laine$^{\rm b}$ 

\vspace{0.8cm}

$^\rmi{a}$%
{\em
Institute of Theoretical Physics, EPFL, 
CH-1015 Lausanne, Switzerland\\}

\vspace*{0.3cm}

$^\rmi{b}$%
{\em
Institute for Theoretical Physics, 
Albert Einstein Center, University of Bern, \\ 
Sidlerstrasse 5, CH-3012 Bern, Switzerland\\}

\vspace*{0.8cm}

\mbox{\bf Abstract}
 
\end{centering}

\vspace*{0.3cm}
 
\noindent
The bulk viscosity of thermalized QCD matter at temperatures above a few
hundred MeV could be significantly influenced by charm quarks because their
contribution arises four perturbative orders before purely gluonic effects. 
In an attempt to clarify the challenges of a lattice study, we determine 
the relevant imaginary-time correlator (of massive scalar densities) 
up to NLO in perturbation theory, and compare with existing data. We find
discrepancies much larger than in the vector channel; this may hint, apart
from the importance of taking a continuum limit, to larger non-perturbative
effects in the scalar channel. We also recall how a transport peak related to
the scalar density spectral function encodes non-perturbative information
concerning the charm quark chemical equilibration rate close to equilibrium.

\vfill

 
\vspace*{1cm}
  
\noindent
October 2013

\vfill

\end{titlepage}

%
\section{Introduction}

Viscosities play an important role in the hydrodynamics 
of finite-size systems, such as those generated 
in heavy ion collision experiments.  
In contrast to thermodynamic functions like the pressure or energy 
density, the dominant contributions to them arise from the slowest
(most weakly interacting) processes relevant for equilibrating energy 
and momentum flows. This may lead to counter-intuitive results; 
for instance, in cosmology, neutrinos or dark matter particles could 
play a dominant role for determining viscosities 
of the cosmic fluid~\cite{sw2,sw1}. 

In this paper we are concerned with the {\em bulk viscosity} of a 
QCD plasma similar to that generated
in heavy ion collision experiments~\cite{adm}. Although less prominent than 
shear viscosity, it also affects the hydrodynamics of 
the system in an interesting way: indeed the bulk viscosity 
could grow rapidly as the temperature decreases below 
the QCD crossover~\cite{ms,ml} and has then been speculated to contribute
to ``clusterization''~\cite{bulk1}--\cite{bulk3} which might 
be viewed as a thermodynamical precursor to the chemical freezeout 
process referred to as hadronization. 

The physical processes relevant for the bulk viscosity are those associated 
with the breaking of scale invariance. 
We suspect that at temperatures $T\gsim 300$~MeV
a significant contribution
may be given by massive quarks, in particular charm quarks. Despite 
enhancement factors~\cite{sommerfeld} the charm quarks are unlikely
to reach chemical equilibrium within the lifetime of heavy ion 
collisions; however it is expected that they depart from
equilibrium on the side of being {\em too many} (cf.\ e.g.\ ref.~\cite{abrs}). 
As suggested by experiments~\cite{star,alice} and 
theoretical determinations~\cite{chm1}--\cite{latt_c}
(similar conclusions have also been reached through phenomenological 
studies~\cite{mt}--\cite{ab} 
as well as investigations in other
gauge theories~\cite{ads1}--\cite{ads3}), 
they also rapidly equilibrate kinetically. 
Nevertheless, to the best of our
knowledge, their contribution to bulk viscosity has not been 
studied in detail,\footnote{%
 In ref.~\cite{adm} massive quarks were considered but 
 results were only worked out for $M \ll \pi T$.
 } 
even though it has been stressed 
that charm quarks may affect a related quantity, 
the speed of sound, in a substantial way~\cite{gt2}.   

In this paper, we do {\em not} address the bulk viscosity per se; rather, we 
consider the imaginary-time correlator from which it can
be extracted non-perturbatively. Our goal is to compute 
a mesonic part of this correlator  (cf.\ \eq\nr{GS_def})
up to next-to-leading order (NLO) in perturbation theory. Given that 
thermal perturbation 
theory in general works better 
for mesonic correlators
than for gluonic ones, and that 
the existence of a non-zero mass scale brings us further
towards the asymptotically free regime, we should 
expect to find quantitatively accurate results at $T\gsim 300$~MeV, as has 
previously been demonstrated in the vector channel~\cite{GVtau}.

The plan of this paper is the following. After elaborating on 
the general physics of the bulk channel (\se\ref{se:background}), 
the setup of the computation is outlined (\se\ref{se:setup})
and the main analytic results are presented (\se\ref{se:formulae}). 
Numerical illustrations and comparisons with non-perturbative data 
comprise \se\ref{se:numerics}, whereas 
\se\ref{se:concl} collects together our findings. Two appendices
contain various details related to the NLO computation. 

%
\section{Physics background}
\la{se:background}

Making use of dimensional regularization (the spacetime dimension
is denoted by $D = 4 - 2\epsilon$) and expressing the QCD action as
\be
 S = \int \! {\rm d}t \int \! {\rm d}^{3-2\epsilon}\vec{x} 
 \, \left\{
  - \frac{1}{4} F^{a\mu\nu}F^a_{\mu\nu} 
  +  \bar{\psi}
 \Bigl( 
   \frac{i}{2} \overleftrightarrow{\!{\bsl{D}}} -  M^{ }_\rmii{B}
 \Bigr)
 \psi
 \right\}
 \;, 
\ee
the trace of the energy-momentum tensor, 
$\hat \theta^{\mu\nu}$, is~\cite{jcc1,jcc2} 
\be
 {\hat{\theta}^{\mu}}_{\;\mu} = c^{ }_\theta\, 
 \underbrace{ \gB^2 F^{a\mu\nu}F^a_{\mu\nu} }_{\equiv\, \theta} \; + \; 
 \bar{\psi} M^{ }_\rmii{B} \psi 
 \;, \la{trace_anomaly}
\ee
where equations of motion were used for the quark fields. 
Here $c_\theta$ is a numerical factor\footnote{%
 More precisely, 
 $
  c^{ }_\theta =  \lim_{\epsilon\to 0 }\frac{D-4}{4g_\rmiii{B}^2} = 
  -\frac{b_0}{2} - \frac{b_1 g^2}{4} + \ldots 
 $, 
 \la{c_theta}
 $
  b_0  =     \frac{11\Nc - 4 T_\rmiii{F}^{ }}{3(4\pi)^2}
 $, 
 $ b_1   =   \frac{34\Nc^2 - 20 \CA^{ } 
   T_\rmiii{F}^{ }-12 C_\rmiii{F}^{ } T_\rmiii{F}^{ } }{3(4\pi)^4}
 $.
 The usual group theory
 factors are  $\Nc = 3$, $\CF \equiv (\Nc^2 - 1)/(2\Nc)$
 and $\TF \equiv \Nf/2$.
 }; 
$\gB^2$ is the bare gauge coupling; and 
$g^2$ 
is a dimensionless renormalized gauge coupling, 
evaluated in the $\msbar$ scheme at the renormalization scale $\bmu$.
The bare mass parameter $M^{ }_\rmii{B}$ is assumed to be 
a diagonal $\Nf\times\Nf$ matrix. 

The trace ${\hat\theta^{\mu}}_{\;\mu}$ 
has a non-zero thermal expectation value, 
\be
 \bigl\langle {\hat\theta^{\mu}}_{\;\mu} \bigr\rangle^{ }_T 
 = e-3p
 = T^5 \frac{{\rm d}}{{\rm d}T}\Bigl(\frac{p}{T^4}\Bigr)
 \;, \la{0pt}
\ee 
where $e$ is the energy density, $p$ is the pressure, 
$T$ is the temperature, and chemical potentials are assumed zero. 
In a scale-invariant theory, in which $p \propto T^4$, 
this expectation value vanishes. 
In QCD, in contrast, it is non-zero, 
both because of dimensional transmutation
and because of non-zero mass parameters. The former
effect originates from loop corrections, and indeed
$e - 3 p = \rmO(c_\theta g^4T^4)$ in massless QCD
at $T \gg 150$~MeV. 
In contrast, in the presence of masses,
the expectation value is non-zero even in a free theory:
\be
 \bigl\langle {\hat\theta^{\mu}}_{\;\mu} \bigr\rangle^{ }_T = 
 4 \Nc \sum_{i=1}^{\Nf} M_i^2 \int_p \frac{\nF{}(E^{ }_{p,i})}{E^{ }_{p,i}} 
 + \rmO(g^2) \;,
 \quad \int_p \equiv \int \! \frac{{\rm d}^3 \vec{p}}{(2\pi)^3}
 \;, \quad  E^{ }_{p,i} \equiv \sqrt{p^2 + M_i^2}
 \;, \la{1pt}
\ee 
where $\nF{}$ is the Fermi distribution and $M_i$ is a renormalized quark
mass of flavour $i$.
The expectation value in \eq\nr{1pt} vanishes in the chiral limit 
$M_i \ll \pi T$ and, because of Boltzmann suppression, also 
for large masses, $M_i \gg \pi T$. Yet it can give a rather 
substantial contribution for $M_i \sim \pi T$; for temperatures
relevant for heavy ion collision experiments, this could be 
the case with charm quarks~\cite{phenEOS}--\cite{buwu}, 
which have a mass $M^{ }_\rmi{c} < M^{ }_{D^0} = 1.86$~GeV.

In the following, we consider 2-point correlation functions 
of ${\hat\theta^{\mu}}_{\;\mu}$. Of particular interest is 
the real-time correlator ($\mathcal{X} \equiv (t,\vec{x})$)
\ba
 \zeta & = & \fr19 
 \lim_{\omega\to 0^+} 
 \biggl\{ 
 \frac{1}{\omega}
 \int_{\mathcal{X}} e^{i \omega t}
 \left\langle 
 \fr12 \left[ {\hat\theta^{\mu}}_{\;\mu} (\mathcal{X}), 
 {\hat\theta^{\mu}}_{\;\mu} (0) \right] 
 \right\rangle^{ }_T
 \biggr\}
 \;, \la{zeta}
\ea
which yields the bulk viscosity. 
Similarly to the expectation value in \eq\nr{0pt}, the bulk viscosity
is perturbatively suppressed in massless QCD~\cite{adm},
\be
  \zeta \sim
  \frac{T^3}{  g^4 } \Bigl(\fr13 - c_s^2\Bigr)^2
  \sim  c_\theta^2\, g^4 T^3
  \;,  \la{zeta_sim}
\ee
where $c_s$ is the speed of sound. In a system with vanishing 
chemical potentials, 
\be
 \fr13 - c_s^2 = \frac{T^3}{3 p''}
 \frac{\rm d}{{\rm d}T}\Bigl(\frac{p'}{T^3}\Bigr)
 \;. \la{ccs}
\ee
Even though different from  
$\langle  {\hat\theta^{\mu}}_{\;\mu} \rangle^{ }_T$
in \eq\nr{0pt}, \eq\nr{ccs} also  
measures the breaking of scale invariance, 
and shows a similar parametric behaviour as
$\langle  {\hat\theta^{\mu}}_{\;\mu} \rangle^{ }_T$. 
Therefore,  we may expect that  
$\zeta$ is significantly influenced by quark masses. 

The argument can be made more precise by relating $\zeta$
to the heavy quark chemical equilibration rate. For $M_i \gg \pi T$, 
${\hat\theta^{\mu}}_{\;\mu}$ of \eq\nr{trace_anomaly} is dominated
by the same term $\bar{\psi} M_i \psi$ that also dominates the heavy
quark Hamiltonian. The shape of the corresponding spectral function
was discussed in ref.~\cite{chemical}; if we take the limit in 
\eq\nr{zeta} all the way down to frequencies 
$\omega\lsim \Gamma^{ }_\rmi{chem}$, where $\Gamma^{ }_\rmi{chem}$
is the heavy quark chemical equilibration rate, then the heavy
quark contribution to the bulk viscosity may be estimated as 
\be
 \delta \zeta 
 \; = \;
   \frac{1}{18T} \lim_{\omega\to 0}
   \biggl\{ \frac{2 M_i^2 \chi^{ }_{\ff}
   \Gamma^{ }_\rmii{chem}}{\omega^2 + \Gamma^2_\rmii{chem} }
   \biggr\}
 \; = \; 
   \frac{M_i^2 \chi^{ }_{\ff}}{9 T \Gamma^{ }_\rmii{chem}}
 \;. \la{G_chem}
\ee
Here $\chi^{ }_{\ff}$ denotes the heavy flavour susceptibility. 
Recalling that at weak coupling~\cite{sommerfeld,GVtau} 
\ba
 \Gamma^{ }_\rmi{chem}
 & = &
 \frac{g^4 \CF}{8\pi M_i^2}
 \Bigl( \Nf + 2 \CF - \fr{\Nc}2 \Bigr)
 \Bigl( \frac{TM_i}{2\pi} \Bigr)^{\fr32}
 e^{-M_i/T}
 \; \biggl( 1 + \rmO\Bigl(\frac{T}{M_i}, 
                         \sqrt{\frac{g^4 M_i}{T}}\Bigr)\biggr)
 \;, \la{G_chem_2}
 \\
 \chi^{ }_{\ff}
 & = &  
 4 \Nc\, \Bigl( \frac{M_iT}{2\pi} \Bigr)^{\fr32} e^{- M_i / T}
 \; \biggl( 1 + \rmO\Bigl(\frac{T}{M_i},\frac{g^2 T}{M_i}\Bigr)\biggr)
 \;,
\ea
it is seen that 
$
 \delta \zeta \sim M_i^4 / g^4 T
$. 
Therefore $\delta\zeta$ 
exceeds the gluonic contribution in \eq\nr{zeta_sim} 
by four perturbative orders, $\rmO(1/g^8)$, 
as is the case also 
for $0 < M_i \ll \pi T$~\cite{adm}.

The purpose of the present study is to consider the imaginary-time 
correlator corresponding to \eq\nr{zeta}:
\be
 \left\langle \int_\vec{x}
 {\hat\theta^{\mu}}_{\;\mu} (X) 
 {\hat\theta^{\mu}}_{\;\mu} (0) 
 \right\rangle^{ }_T
 \;,
\ee 
where $X \equiv (\tau,\vec{x})$, 
$0 < \tau < 1/T$, and heavy quarks are assumed to be in full
equilibrium.\footnote{%
 Given that in real-world heavy ion collisions
 heavy quarks appear in overabundance,
 this can in some sense be considered a conservative assumption. 
 } 
Recalling \eq\nr{trace_anomaly}, 
this correlator contains three terms. The 2-point correlator of 
$c_\theta\, \theta$ has
been computed up to NLO in the 
weak-coupling expansion~\cite{Bulk_OPE,Bulk_wdep}
and compared with high-precision lattice simulations~\cite{hbm}
(the lattice data contain no continuum extrapolation but are
augmented by a tree-level improvement). Here we 
consider the 2-point correlator of 
the fermionic part, cf.\ \eq\nr{GS_def}, 
and compare with quenched lattice data from ref.~\cite{ding2}
(these data contain neither a continuum extrapolation nor 
tree-level improvement but have a fairly fine lattice spacing).
In addition, 
independently of numerical data, we estimate within perturbation 
theory whether the charm contribution could dominate
over the purely gluonic one at phenomenologically
interesting temperatures.

Returning to bulk viscosity, it should be noted that 
in a practical heavy ion collision the system has a finite (short)
lifetime, which implies that there is a typical frequency
$\omega\sim {\rm fm}/c$ that can play a role in the hydrodynamical
evolution of the system. If the width of the charm quark transport
peak is narrower than this ($\Gamma^{ }_\rmi{chem} < \omega$), 
then only gluons and light quarks contribute to
the bulk viscosity.  If $\Gamma^{ }_\rmi{chem} \sim \omega$,
then heavy quarks should be included but to do this consistently
requires going beyond a hydrodynamical description, perhaps by
employing kinetic theory.  If $\Gamma^{ }_\rmi{chem} > \omega$,
a hydrodynamical description applies and the heavy quark contribution
should be added to the bulk viscosity. In each case the chemical
equilibration rate $\Gamma^{ }_\rmi{chem}$ is seen to be a fundamental
quantity, whose non-perturbative determination as a function of 
the heavy quark mass would be more than welcome.

%
\section{Setup of the computation}
\la{se:setup}

For simplicity, we consider a situation in the following in which there
is {\em one} quenched heavy quark, of bare mass $M^{ }_\rmii{B}$ and 
renormalized mass $M$
(different schemes are specified presently), and $\Nf$  
massless dynamical quarks. (This means that from now on the 
quarks in \se\ref{se:background} should be thought of as having
$\Nf + 1$ flavours.)
The imaginary-time correlator considered is 
\ba
 G^{ }_\rmii{S}(\tau)
 & \equiv & 
 M_\rmii{B}^2 \int_\vec{x} 
 \Bigl\langle 
 (\bar\psi \psi) (\tau,\vec{x}) 
 \;
 (\bar\psi \psi) (0,\vec{0})
 \Bigr\rangle^{ }_T
 \;, \quad
 0 < \tau < \beta 
 \;, \quad \beta \equiv \frac{1}{T}
 \;, \la{GS_def}
\ea
where $\psi$ denotes a single-flavour heavy quark Dirac spinor. 
Defined this way, the NLO scalar density correlator is finite 
after mass and gauge coupling renormalization. 

We compute the correlator by first determining the corresponding correlator
in momentum space, with an external four-momentum 
\be
 Q \equiv (\omega_n,\vec{0}) \;,
\ee
where $\omega_n$ is a bosonic Matsubara frequency. Denoting that result
by $\tilde G^{ }_\rmii{S}(\omega_n)$, the 
correlator of \eq\nr{GS_def} is obtained from 
\be
 G^{ }_\rmii{S}(\tau) = T \sum_{\omega_n} e^{-i \omega_n \tau}
 \, 
 \tilde G^{ }_\rmii{S}(\omega_n)
 \;. 
\ee

Given that the definition in  
\eq\nr{GS_def} involves a bare parameter, the issue of renormalization
and quark mass definitions plays an important role. 
The bare mass parameter, $M_\rmii{B}^2$, can be expressed as 
$ M_\rmii{B}^2  =   M^2 + \delta M^2$, where the choice of $\delta M^2$
defines a scheme. We write 
\ba
 \delta M^2 & = & 
 - \frac{6 g^2 \CF M^2}{(4\pi)^2} 
 \biggl( \frac{1}{\epsilon} + \ln\frac{\bmu^2}{M^2} + \fr43 + \delta \biggr) 
 + \rmO(g^4)
 \;,  \la{MB}
\ea
where $\bmu$ is 
the scale parameter of the $\msbar$ scheme, 
and terms of $\rmO(\epsilon)$ were omitted. 
For $\delta = 0$, $M^2$ corresponds to a pole mass, 
which tends to compactify analytic expressions but
is ambiguous on the non-perturbative level and also leads to 
problems of convergence (see below). 
If we choose 
\be
 \delta =  - \ln\frac{\bmu^2}{M^2} - \fr43
 \;, \la{m_msbar}
\ee
then $M^2$ stands for the $\msbar$ mass, which we denote
by $m^2(\bmu)$; its asymptotic running reads
\be
 m(\bmu) = m(\bmu_\rmi{ref})
 \biggl[
   \frac{\ln(\bmu_\rmi{ref}/\Lambdamsbar)}{\ln(\bmu/\Lambdamsbar)} 
 \biggr]^{\frac{9 \CF}{11\CA - 4 \TF}}
 \;, \la{m_bmu}
\ee
where typically 
$\bmu_\rmi{ref} \equiv 2$~GeV is chosen~\cite{pdg}. 
We leave the value of $\delta$ open for the moment. 

%
\section{Analytic results}
\la{se:formulae}

%
\subsection{Wick contractions}

Denoting propagators by 
\be
 \Delta^{ }_P \equiv P^2 + M^2
 \; 
\ee
and Matsubara sum-integrals by 
$\Tinti{ \{ P \} } \equiv T\sum_{ \{ p_n \} } \int_\vec{p}$, 
where $\{ P \}$ stands for fermionic Matsubara momenta,
the tree-level correlator reads
\ba
 \nn[-10mm]
 \ConnectedC(\TAsc,\TAsc) 
 & = & 
 - 2 \CA M^2 \Tint{\{P\}}
 \biggl\{
 -   \frac{2}{\Delta^{ }_{P} } 
 +  \frac{Q^2 + 4 M^2}{\Delta^{ }_{P}\Delta^{ }_{P-Q} }
 \biggr\}
 \;. \la{GS_lo} \\[-10mm] \nonumber
\ea
The counterterm contribution is 
\ba
 \nn[-10mm] 
 \ConnectedD(\TAsc,\TAsc) 
 & = & -2 \CA \delta M^2 
 \Tint{\{P\}}
 \biggl\{ 
 - \frac{2}{\Delta^{ }_{P}}
 + \frac{2M^2}{\Delta^2_{P} } 
 + \frac{Q^2 + 8 M^2} {\Delta^{ }_{P} \Delta^{ }_{P-Q} }
 - \frac{2M^2 (Q^2 + 4 M^2)}{\Delta^2_{P}\Delta^{ }_{P-Q} }
 \biggr\}
 \;, \hspace*{5mm} \la{GS_ct}
\ea
whereas the ``genuine'' 2-loop graphs amount to
\ba
 \nn[-10mm]
 && \hspace*{-1.5cm}
 \ConnectedA(\TAsc,\TAsc,\TAgl) \; + 
 \ConnectedB(\TAsc,\TAsc,\TLgl) \quad
 = \, 4 g^2 \CA \CF M^2 \Tint{K\{P\}} 
 \biggl\{  \nonumber  \\[-5mm] 
 &&  \hspace*{-1cm}
 - \, \frac{D-2}{K^2\Delta^2_{P} } + 
 \frac{D-2}{\Delta^2_{P} \Delta^{ }_{P-K} } +
 \frac{2}{K^2\Delta^{ }_{P}\Delta^{ }_{P-K} } -
 \frac{4 M^2}{K^2\Delta^2_{P}\Delta^{ }_{P-K} }  
 \nn
 && \hspace*{-1cm}
 + \, \frac{(D-2)(Q^2 + 4 M^2)}{K^2 \Delta^2_{P} \Delta^{ }_{P-Q} }
 - \frac{(D-2)(Q^2 + 4 M^2)}
        {\Delta^2_{P} \Delta^{ }_{P-K}\Delta^{ }_{P-Q} }
 \nn
 &&  \hspace*{-1cm}
 - \, \frac{16 M^2 - 2 (D-2) K\cdot Q + 4 Q^2}
 {K^2 \Delta^{ }_{P} \Delta^{ }_{P-K} \Delta^{ }_{P-Q} } 
 +  \frac{4 M^2 (Q^2 +  4  M^2) }
 {K^2 \Delta^2_{P} \Delta^{ }_{P-K} \Delta^{ }_{P-Q} } 
 \nn
 &&  \hspace*{-1cm} 
 - \, \frac{2(D-2) M^2 + \fr12 (D-4) Q^2 }
 {\Delta^{ }_{P} \Delta^{ }_{P-K} \Delta^{ }_{P-Q} \Delta^{ }_{P-K-Q}}
 + \frac{8 M^4 + 6 M^2 Q^2 + Q^4}
 {K^2 \Delta^{ }_{P} \Delta^{ }_{P-K}\Delta^{ }_{P-Q} \Delta^{ }_{P-K-Q} }
 \biggr\}
 \;. \la{GS_nlo}
\ea

%
\subsection{Result after Matsubara sums and angular and partial integrals}

All the Matsubara sums (cf.\ appendix~A)
as well as some 
of the angular integrals appearing 
in \eqs\nr{GS_lo}--\nr{GS_nlo}
can be carried out analytically; 
in addition partial integrations
permit for simplifications
(cf.\ ref.~\cite{ibp} for the massless case).  
To display the results, we employ the functions 
\be
 D_{E_1 \cdots E_k}^{E_{k+1} \cdots E_n}(\tau) 
 \equiv
 \frac{
 e^{ (E_1 + \cdots + E_k)(\beta - \tau) +
     (E_{k+1} + \cdots + E_n)\tau } 
   + 
 e^{ (E_1 + \cdots + E_k)\tau +
 (E_{k+1} + \cdots + E_n)(\beta - \tau) } 
 }{[e^{\beta E_1} \pm 1] \cdots [e^{\beta E_n} \pm 1]}
 \;, \la{D_def}
\ee
where the sign in the denominator
is chosen according to whether the 
particle is a boson or a fermion.  
The energy variables 
\be
 \epsilon_k \equiv |\vec{k}|
 \;,  \quad 
 E_p \equiv \sqrt{p^2 + M^2}
 \;, \quad
 E_{pk} \equiv \sqrt{(\vec{p-k})^2 + M^2}
 \;,  \quad
 E_{pk}^\pm \equiv \sqrt{(p\pm k)^2 + M^2}
 \la{energies}
\ee
appear frequently, 
and we denote $D^{ }_{2E_p} \equiv D^{ }_{E_pE_p}$.
Then the leading-order (LO) result reads
\ba
 \left. 
   G_\rmii{S}^\rmii{LO}(\tau) 
 \right|_\rmii{$\tau$-dep.}
 & = & 
 2 \CA M^2 
 \int_{p} 
   \frac{p^2 D^{ }_{2E_p}(\tau) }{E_p^2}  
  \;, \la{GSLOtaul}
 \\
 \left. 
   G_\rmii{S}^\rmii{LO}(\tau) 
 \right|_\rmii{const.}
 & = &  - 4 \CA M^2 
 \int_{p} 
 \frac{M^2 T \nF{}'(E_p)}{E_p^2} 
  \;, \la{GSLOconst}
\ea
where a $\tau$-independent part stemming from an approximate
transport peak in the corresponding spectral function has been 
separated. Scheme dependence can be expressed as 
\ba
 \left. 
   \frac{ 
     \Delta^{ }_\delta G_\rmii{S}^\rmii{NLO}(\tau) 
  }{ 
  4 g^2 \CA \CF M^2
  } 
 \right|_\rmii{$\tau$-dep.}
 & = & 
 \frac{3 \delta}{32 \pi^2}
 \int_{p} \, D^{ }_{2E_p}(\tau)  
  \biggl( \frac{5 M^2 }{E_p^2}  - 2 \biggr)  
  \;, \la{GS_LO_tau_delta}
 \\
 \left. 
  \frac{ 
   \Delta^{ }_\delta G_\rmii{S}^\rmii{NLO}(\tau) 
  }{ 
  4 g^2 \CA \CF M^2
  } 
 \right|_\rmii{const.}
 & = &  
 - \frac{3 \delta}{16 \pi^2}
 \int_{p} 
 \frac{M^2 T \nF{}'(E_p)}{p^2} 
  \biggl( \frac{5 M^2 }{E_p^2}  - 4 \biggr)  
  \;. \la{GS_LO_const_delta}
\ea
Introducing the shorthand notations 
\be
 \Delta_{\sigma\tau} \equiv \epsilon_k + \sigma E_p + \tau E_{pk}
 \;, \quad
 \Delta_{\sigma} = E_p + \sigma E_{pk} 
 \;, \la{def_Delta}
\ee
the $\tau$-dependent part of the NLO correction reads
(after renormalization, cf.\ appendix~B)
\ba
 & & \hspace*{-1cm}
 \frac{ 
  \left. G_\rmii{S}^\rmii{NLO} \right|_\rmii{$\tau$-dep.}
  }{ 
  4 g^2 \CA \CF M^2
  } = 
 \int_{p}  
 \frac{D^{ }_{2E_p}(\tau)}{8\pi^2} \biggl[ 
 \frac{M^2}{E_p^2} 
 \biggl( 1 + \frac{p}{E_p} \ln \frac{E_p + p}{E_p - p}\biggr)
 - 1
 - \biggl(1 - \frac{M^2}{E_p^2} \biggr) 
   \int_0^\infty \!\!\! {\rm d}k\, \frac{2 \theta(k)}{k}
\biggr] 
  \la{GS_NLO_tau_final} \\ 
 & + & \!\!  \int_{p,k} \!\! \mathbbm{P} \Biggl\{   
 \int_{z} 
 \frac{D^{ }_{\epsilon_k E_p E_{pk}}(\tau) M^2}{\epsilon_k E_p E_{pk}
 \Delta_{+-}\Delta_{-+} }
 \biggl[ 
  \frac{\epsilon_k^2 + (E_p + E_{pk})^2 }{4 M^2}
  + 
  \frac{\Delta^{ }_{--}}{\Delta^{ }_{++}}
 - \frac{\epsilon_k^2}{\Delta^{ }_{+-}\Delta^{ }_{-+}}
 + \frac{4\epsilon_k^2 M^2}{\Delta_{++}^2 \Delta^{ }_{+-}\Delta^{ }_{-+}}
 \biggr]
 \nn 
 &  &  \quad + \, 
 \int_{z} 
 \frac{D^{\epsilon_k}_{E_p E_{pk}}(\tau) M^2}{\epsilon_k E_p E_{pk}
 \Delta_{+-}\Delta_{-+}
 }
 \biggl[ 
  \frac{\epsilon_k^2 +  (E_p +  E_{pk})^2}{4M^2}
  + 
  \frac{\Delta^{ }_{++}}{\Delta^{ }_{--}}
 - \frac{\epsilon_k^2}{\Delta^{ }_{+-}\Delta^{ }_{-+}}
 + \frac{4\epsilon_k^2 M^2}{\Delta_{--}^2 \Delta^{ }_{+-}\Delta^{ }_{-+}}
 \biggr]
 \nn 
 &  & \quad - \, 
 \int_{z} 
 \frac{2 D^{E_p}_{\epsilon_k E_{pk}}(\tau) M^2}{\epsilon_k E_p E_{pk}
  \Delta_{++}\Delta_{--}
 }
 \biggl[ 
   \frac{\epsilon_k^2 +  (E_p -  E_{pk})^2}{4M^2}
  +  
  \frac{\Delta^{ }_{+-}}{\Delta^{ }_{-+}}
 - \frac{\epsilon_k^2}{\Delta^{ }_{++}\Delta^{ }_{--}}
 + \frac{4\epsilon_k^2 M^2}{\Delta_{-+}^2 \Delta^{ }_{++}\Delta^{ }_{--}}
 \biggr]
 \nn 
 & + &  \frac{ D^{ }_{2E_p}(\tau)}{ 2\epsilon_k^3} 
 \biggl(1  - \frac{M^2}{E_p^2} \biggr) 
  \, \biggl[ \;
 1 - \frac{
    E_p^2 (E_{pk}^+ - E_{pk}^-)
     -
    p\epsilon_k (E_{pk}^+ + E_{pk}^-)
    }{2p(E_p^2 - \epsilon_k^2)}
 - \frac{\epsilon_k^2 M^2 (E_{pk}^+ - E_{pk}^-) }
    {p (E_p^2 - \epsilon_k^2) E_{pk}^+  E_{pk}^- }
 \nn &   & 
 \quad - \,
 \frac{ 2 E_p^2  - M^2 }{2p E_p}
 \biggl(  
     \ln\biggl| \frac{(E_p + p)(2p-\epsilon_k)}
                     {(E_p - p)(2p+\epsilon_k)}
        \biggr|
    +\ln\biggl| \frac{1 - {\epsilon_k^2} / {(E_p + E_{pk}^-)^2}}
                 {1 - {\epsilon_k^2} / {(E_p + E_{pk}^+)^2}}
        \biggr|
 \biggr) \;
 + \theta(k) 
 \biggr]
 \nn 
 &  + &
 \frac{D^{ }_{2E_p}(\tau)  \nB{}(\epsilon_k)}{\epsilon_k}
  \biggl[ \;
     -  
     \frac{1}{2 p E_p} \ln\frac{E_p + p}{E_p - p}
  \nn & & \quad  + \, 
   \biggl( 1 - \frac{M^2}{E_p^2} \biggr)
   \biggl( \frac{1}{\epsilon_k^2} - 
  \frac{1}{2 p^2} -  \frac{ 2 E_p^2  - M^2 }{2 p E_p \epsilon_k^2}
   \ln\frac{E_p + p}{E_p - p}
   \biggr) 
 \; \biggr]
 \nn &  + &    
 \frac{D^{ }_{2E_p}(\tau)  \nF{}(E_{k})}{E_{k}} \, 
 \biggl[ \;
 - \frac{1}{2E_p^2}
 - \frac{2 E_p^2 + M^2}{2({p^2-k^2})E_p^2}
 + \frac{M^2}{p k E_p^2} \ln \biggl| \frac{p+k}{p-k} \biggr|
 \nn &   &  \quad + \, 
 \frac
 {E_p^2(E_p^2 + E_k^2 - 3 M^2) + M^4}
 {2 p k (E_p - E_k) E_p^3}
 \biggl( 
   \ln \biggl| \frac{p+k}{p-k} \biggr| + 
  \frac{E_k}{E_p + E_k} 
   \ln \biggl| \frac{M^2 + E_p E_k + p k}{M^2 + E_p E_k - p k} \biggr|
 \biggr)
 \biggr]
 \Biggr\} 
 \;,\nonumber 
\ea
where $\mathbbm{P}$ denotes a principal value, 
$\int_z$ an integral over the angles between $\vec{p}$ and $\vec{k}$
(with $\int_z 1 = 1$), 
$\nB{}$ the Bose distribution, and the function
$\theta(k)$ is specified in \eq\nr{theta_k}.
The constant contribution reads 
\ba
 & & \hspace*{-1cm}
 \frac{ 
  \left. G_\rmii{S}^\rmii{NLO} \right|_\rmii{const.}
  }{ 
  4 g^2 \CA \CF M^2
  } 
  =  \int_{p}  T \nF{}'(E_p)   \frac{M^2}{E_p^2}  
  \biggl\{ 
  \frac{3 }{4 \pi^2 } 
  \; + \; 
  \int_{k} \frac{\nB{}(\epsilon_k)}{\epsilon_k} \biggl[ 
   \frac{1}{p^2}
 \biggr]
 \nn & + &  
 \int_{k} \frac{\nF{}(E_{k})}{E_{k}} \biggl[ 
   \frac{1}{p^2}\biggl(1 - \frac{M^2}{k^2}  \biggr) 
  + \frac{1}{k^2} + \frac{1}{E_k^2} 
  + \frac{4 E_k^2 - M^2}{2 p k E_k^2}
    \ln \biggl| \frac{p+k}{p-k} \biggr|  
 \biggr]
 \biggr\} \;. \hspace*{8mm}
  \la{GS_NLO_const_final}
\ea
Note that for $M \gg \pi T$ the leading thermal correction
in \eq\nr{GS_NLO_const_final}, from 
$ 
 \int_{k} \nB{}(\epsilon_k) /  \epsilon_k 
$, 
agrees with the leading NLO thermal correction to 
$\chi^{ }_{\ff}$~\cite{GVtau}, as is to be expected from 
integration over the transport peak in \eq\nr{G_chem}
with the standard relation in \eq\nr{GS_relation}. 

%
\subsection{Infrared and ultraviolet regimes}

For the numerical evaluation of \eq\nr{GS_NLO_tau_final}
it must be kept in mind that individual
parts of the expression contain divergences; only the 
sum is well-defined. In particular,
at small $k$ there is a divergence originating from  
the terms integrated over $z$ in \eq\nr{GS_NLO_tau_final} 
and from terms where the integral had already been carried out. 
For the latter type the small-$k$ part reads
\be
 \frac{ D^{ }_{2E_p}(\tau)}{ \epsilon_k^3} 
 \biggl(1  - \frac{M^2}{E_p^2} \biggr) 
  \, \biggl( \;
 1 - 
 \frac{ 2 E_p^2  - M^2 }{2p E_p}
     \ln\frac{E_p + p}
                     {E_p - p}
 \;
 \biggr)
 \biggl[ \; 
   \fr12 + \nB{}(\epsilon_k)
 \, \biggr]
 \;,
\ee
containing both vacuum and Bose-enhanced structures, leading
to logarithmic and powerlike divergences. When terms 
of both origins are summed together, the small-$k$ divergences cancel.  

At large  $k$ the leading asymptotic behaviour is 
\be
 \frac{ D^{ }_{2E_p}(\tau)}{ \epsilon_k^3} 
 \biggl(1  - \frac{M^2}{E_p^2} \biggr) 
  \, \biggl( \;
  \frac{ 3 E_p^2}{ 2 \epsilon_k^2 } 
 \biggr)
 \;. 
\ee
Although integrable this expression is only power-suppressed, 
and for numerical handling it may be advantageous to 
accelerate the decrease. The leading tail 
can be subtracted with the help of the auxiliary function $\theta(k)$
appearing on the 1st and 6th rows of \eq\nr{GS_NLO_tau_final}, 
for instance 
\be
 \theta(k) 
 \equiv
 - \frac{3 E_p^2 \Theta(k-k_\rmii{min})}{k^2 + \lambda^2}
 \;, \quad
  \int_0^\infty \!\! {\rm d}k\, \frac{\theta(k)}{k}
 = 
  - \frac{3 E_p^2  }{2\lambda^2}
  \, \ln \Bigl( 1 + \frac{\lambda^2}{k_\rmii{min}^2} \Bigr)
 \;. \la{theta_k}
\ee 
The resolution obtained can be increased with a suitable tuning
of $\lambda$ and $k_\rmi{min}$; as typical values we have 
used $\lambda \sim M/100$, $k_\rmi{min} \sim 10 T$.

%
\section{Numerical evaluations}
\la{se:numerics}

%
\subsection{Parameter choices}

For a comparison with lattice simulations 
certain parameter values need to be fixed. 
Both the gauge coupling and the masses are running parameters; 
in accordance with ref.~\cite{ding2} we set $\Lambdamsbar \simeq 216$~MeV
to fix the gauge coupling 
(note that the simulations in ref.~\cite{ding2} are for $\Nf = 0$).
As a rule the gauge coupling is evolved with 3-loop running
(cf.\ ref.~\cite{rit1} and references therein), 
with some exceptions as specified below.  
For the running masses a standard choice according to ref.~\cite{pdg}
is to evaluate them at $\bmu^{ }_\rmi{ref} = 2$~GeV, and then 
the charm mass is $m_c(\bmu^{ }_\rmi{ref}) = 1.275(25)$~GeV. 
However in ref.~\cite{ding2} the simulations correspond to 
$m_c(m_c) = 1.094(1)$~GeV, which according to 4-loop evolution~\cite{rit2}
as employed in ref.~\cite{ding2} 
corresponds to $m_c(\bmu^{ }_\rmi{ref}) = 0.967(1)$~GeV.   
(The numbers cited should probably be assigned generous
systematic uncertainties.)

As usual, the perturbative predictions display a residual dependence
on the $\msbar$ scale $\bmu$, which can be 
used for estimating the uncertainties of a fixed-order computation. 
We always evaluate the result at some ``optimal'' scale $\bmu_\rmi{opt}$
as well as at $0.5\, \bmu_\rmi{opt}$ and $2.0\, \bmu_\rmi{opt}$; the 
maxima and minima among these three numbers are then used for constructing
``error bands'' for the perturbative predictions. Obviously such 
bands only serve as lower bounds for the systematic uncertainties
related to the perturbative computation. 

In order to display the results in a useful way, 
they should be normalized to an appropriate expression. 
We have considered three options for this purpose:  
a free result with a light quark mass ($M \ll \pi T$); 
the scalar correlator in pure Yang-Mills theory; as well as 
a ``reconstructed'' correlator following from a zero-temperature
spectral function.

%
\subsection{Normalization to a free correlator}

The LO result for the scalar channel correlator
is given in \eqs\nr{GSLOtaul}, \nr{GSLOconst}.
It vanishes in the chiral limit, and cannot be evaluated 
in a closed form for a general mass. Nevertheless, if we assume that 
$M \ll \pi T$, we can set the mass to zero within the integrand, 
yielding~\cite{ff}
\ba
 G_\rmii{S}^\rmi{naive}(\tau) & \equiv &   
 \Nc T^3 M^2 \biggl[ 
  \pi \, (1-2\tau T) \,
  \frac{1 + \cos^2(2\pi\tau T)}{\sin^3(2\pi \tau T)}
 + 
 \frac{2 \cos(2\pi \tau T)}{\sin^2(2\pi \tau T)}
  \biggr]
 \;. \la{GSfree_naive}
\ea
It turns out, however, that this expression does not compare well
with lattice data even for $\tau \ll 1/T$; the situation is 
illustrated in \fig\ref{fig:GSnaive}. The reason can be understood
as being related to running effects, as we now explain. 

\begin{figure*}[tb]


\centerline{%
 \epsfysize=8.5cm\epsfbox{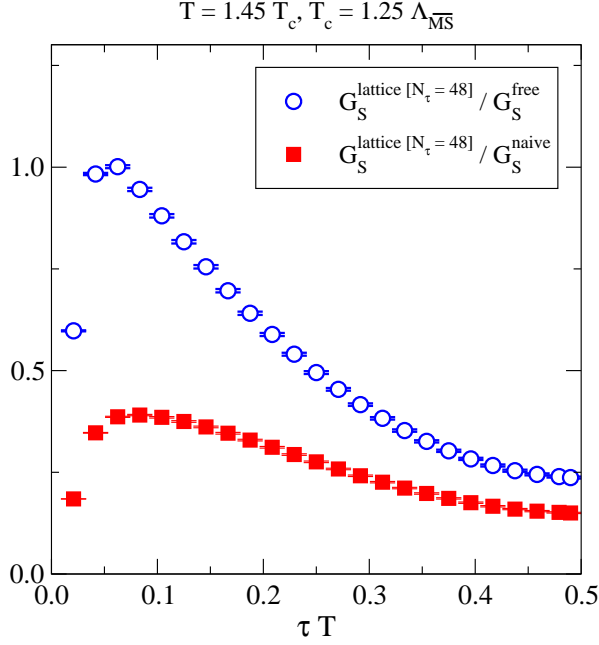}
}

\caption[a]{\small
The ratio 
$G_\rmii{S}^\rmi{lattice} / G_\rmii{S}^\rmi{free}$
(with $m(\bmu_\rmi{ref}) = 967$~MeV; open symbols) 
as well as
$G_\rmii{S}^\rmi{lattice} / G_\rmii{S}^\rmi{naive}$
(with $M/T = 3.6$ as extracted from a comparison of numerical data
and NLO expressions in the vector channel~\cite{GVtau}; 
closed symbols).
The lattice data are from ref.~\cite{ding2} and contain 
no continuum extrapolation.  
It is seen that normalization to $G_\rmii{S}^\rmi{free}$ 
(cf.\ \eq\nr{GSfree})
yields results
closer to unity at short distances; the very smallest 
distances are affected by lattice artifacts. 
}

\la{fig:GSnaive}
\end{figure*}

Let us start by considering
\eq\nr{GSfree_naive} at short distances, 
\be
 G^\rmi{naive}_\rmii{S}(\tau) 
 \;\; \stackrel{\tau \ll  \frac{1}{T}  }{\approx} \;\;
 \frac{\CA M^2}{4\pi^2 \tau^3}
 \;. 
\ee
Then compute the NLO correction to this result; 
making use of the spectral function in \eq\nr{full_vacS_asympt}, 
the LO + NLO expression reads
\be
 G^\rmii{LO+NLO}_\rmii{S}(\tau) 
 \;\; \stackrel{\tau \ll \frac{1}{M}, \frac{1}{T}  }{\approx} \;\;
 \frac{\CA M^2}{4\pi^2 \tau^3}
 \biggl[ 1   
 + \frac{6 g^2 \CF }{(4\pi)^2}
 \biggl(
  \ln{\tau^2 M^2} - \delta - \fr32 + 2 \gammaE  
 \biggr) 
 \biggr]
 \;. \la{GS_asympt}
\ee
We note that if $M^2$ is kept fixed, 
$G^\rmii{LO+NLO}_\rmii{S}$ turns negative
at small $\tau$, but 
that simultaneously the loop expansion breaks down because
the NLO correction overtakes the LO term. 
The problem can be rectified if we use running 
parameters: going to the $\msbar$ scheme; 
choosing $\delta$ according to \eq\nr{m_msbar} and subsequently replacing
$
 M^2 \to m^2(\bmu) 
$
according to \eq\nr{m_bmu};
and taking $\bmu$ to scale with $\tau$ according to 
a ``fastest apparent convergence'' criterion,    
\be
 \bmu\to \frac{\beta e^{\frac{1}{12} - \gammaE}}{\tau(\beta-\tau)}
 \;, \la{bmu_asympt}
\ee 
such that the NLO correction in \eq\nr{GS_asympt} always 
remains small, we are led to define
\ba
 G_\rmii{S}^\rmi{free}(\tau) & \equiv &   
 \Nc T^3  
 m^2(\bmu_\rmi{ref})
 \Biggl\{ 
   \frac{\ln\bigl[ \frac{\bmu_\rmi{ref}}{\Lambdamsbar} \bigr] }
        {\ln\bigl[ 
 \frac{\beta e^{\frac{1}{12} - \gammaE} }
   {\tau(\beta-\tau )\Lambdamsbar} \bigr]} 
 \Biggr\} ^{\frac{18 \CF}{11\CA - 4 \TF}}
  \nn & & \times \, 
 \biggl[ 
  \pi \, (1-2\tau T) \,
  \frac{1 + \cos^2(2\pi\tau T)}{\sin^3(2\pi \tau T)}
 + 
 \frac{2 \cos(2\pi \tau T)}{\sin^2(2\pi \tau T)}
  \biggr]
 \;. \la{GSfree}
\ea
Let us stress that this is a {\em definition} rather than an exact 
result; indeed only 
the asymptotics at $\tau \ll \beta$ can be fixed unambiguously thanks 
to asymptotic freedom. Nevertheless, as seen in \fig\ref{fig:GSnaive}, 
\eq\nr{GSfree} agrees much better with lattice data at small distances
than \eq\nr{GSfree_naive}. 

\begin{figure*}[tb]


\centerline{%
 \epsfysize=8.5cm\epsfbox{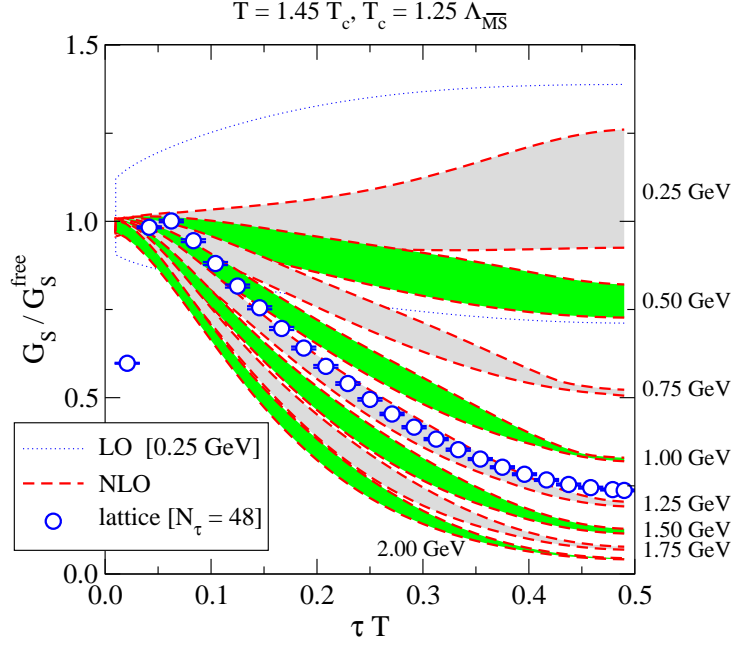}
}

\caption[a]{\small
The ratio $G_\rmii{S} / G_\rmii{S}^\rmi{free}$ 
(cf.\ \eq\nr{GSfree})
as a function of $\tau T$
and the quark mass $m(\bmu_\rmi{ref})$, indicated next to the coloured
bands. The bands reflect the uncertainty related to the choice of 
the renormalization scale as specified in the text.  
The lattice simulations are from ref.~\cite{ding2} and correspond to 
$m(\bmu_\rmi{ref}) \simeq 967$~MeV.  The data at $\tau T > 0.15$ 
are suppressed with respect to the continuum prediction.
}

\la{fig:GSfree}
\end{figure*}

The full NLO 
results normalized to \eq\nr{GSfree} are shown in \fig\ref{fig:GSfree}
as a function of $m(\bmu_\rmi{ref})$. 
As illustrated with the example of $m(\bmu_\rmi{ref}) = 0.25$~GeV, the 
renormalization scale dependence gets significantly reduced when going
from the LO to the NLO level. The NLO results for 
$m(\bmu_\rmi{ref}) = 1$~GeV agree with the lattice data 
at small $\tau$, but at large $\tau$ a discrepancy sets in. 
We return to the discrepancy in connection with \fig\ref{fig:GStheta}.

%
\subsection{Comparison with gluonic effects}

The spectral function related to the gluonic part of the trace anomaly, 
i.e.\ the operator $c_\theta \, \theta$ defined in \eq\nr{trace_anomaly},
was computed up to NLO in ref.~\cite{Bulk_wdep}. At LO it reads
(for $\omega \gsim \pi T$)
\be
 \rho_\theta^\rmii{LO}(\omega) = 
 \frac{\CA \CF\, c_\theta^2\, g^4 \omega^4}{2\pi} 
 \biggl[
   1 + 2 \nB{}\Bigl( \frac{\omega}{2} \Bigr) 
 \biggr]
 \;, \la{rho_theta_lo}
\ee
which yields the correlator 
\be
 \frac{ G_{\theta}^\rmi{naive}(\tau) }{ c_\theta^2\, g^4 \CA \CF}  \equiv    
 128 \pi^2 T^5  \biggl[ 
  \pi \, (1-2\tau T) \,  
  \frac{2\cos(2\pi\tau T) + \cos^3(2\pi\tau T)}{\sin^5(2\pi \tau T)}
 + 
 \frac{1 + 2 \cos^2(2\pi \tau T)}{\sin^4(2\pi \tau T)}
  \biggr]
 \;. \la{Gthetafree_naive}
\ee
However, there are again loop corrections which 
imply that the running of the coupling needs to be taken into account; 
the asymptotics reads~\cite{hbm_old}
\be
 \rho^\rmi{vac}_{\theta}(\omega) 
 \;\; \stackrel{\omega \gg \pi T}{\approx} \;\;
 \frac{\Nc \CF\, c_\theta^2\, \omega^4 }{2\pi}
 \biggl\{ 
   g^4(\bmu) + \frac{g^6(\bmu)\Nc}{(4\pi)^2}
  \biggl(
   \frac{22}{3} \ln\frac{\bmu^2}{\omega^2}  + \frac{73}{3} 
  \biggr)
 + \rmO(g^8) 
 \biggr\}
 \;,
\ee 
with thermal corrections strongly suppressed at 
$\omega \gg \pi T$~\cite{simon,Bulk_OPE}.
Inserting into \eq\nr{GS_relation} it is seen that 
at $\tau \ll \beta$
the effects from the running can be captured
through a choice analogous to \eq\nr{bmu_asympt}, {\em viz.} 
\be 
 \bmu\to \frac{\beta e^{\frac{14}{33} - \gammaE}}{\tau(\beta-\tau)}
 \;, \la{bmu_asympt_2}
\ee
and this leads us to define 
\ba
 G_{\theta}^\rmi{free}(\tau) & \equiv &   
 2 c_\theta^2 \CA \CF T^5  
 \Biggl\{ 
   \frac{192 \pi^3 }
        {(11 \CA - 4 \TF)
       \ln\bigl[ \frac{\beta e^{\frac{14}{33} - \gammaE}}
      {\tau(\beta-\tau )\Lambdamsbar} \bigr]} 
 \Biggr\}^{2}
 \nn & & \times \, 
 \biggl[ 
  \pi \, (1-2\tau T) \, 
  \frac{2\cos(2\pi\tau T)  + \cos^3(2\pi\tau T)}{\sin^5(2\pi \tau T)}
 + 
 \frac{1 + 2 \cos^2(2\pi \tau T)}{\sin^4(2\pi \tau T)}
  \biggr]
 \;. \la{Gthetafree}
\ea
Only the asymptotics at $\tau \ll \beta$ is unambiguously fixed, 
otherwise \eq\nr{Gthetafree} represents a choice. 
(To be concrete, for this definition we have evaluated the coupling 
appearing in $c_\theta$, cf.\ footnote~\ref{c_theta}, at 1-loop level 
at the scale indicated by \eq\nr{bmu_asympt_2}.)

\begin{figure*}[tb]


\centerline{%
 \epsfysize=8.5cm\epsfbox{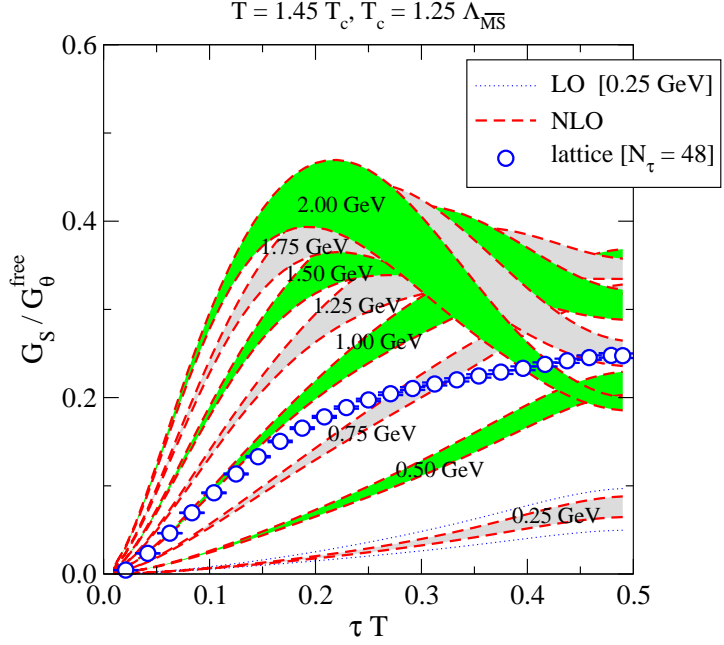}
}

\caption[a]{\small
The ratio $G_\rmii{S} / G_{\theta}^\rmi{free}$ 
(cf.\ \eq\nr{Gthetafree}) as a function of $\tau T$
and the quark mass $m(\bmu_\rmi{ref})$, indicated next to the coloured
bands. The bands reflect the uncertainty related to the choice of 
the renormalization scale. 
Recalling $m_c(\bmu^{ }_\rmi{ref}) = 1.275(25)$~GeV~\cite{pdg} 
it is seen that at large distances the charm quark contribution 
to the bulk channel correlator could be as large as $30 - 40$\%; 
however lattice data~\cite{ding2} at $m(\bmu^{ }_\rmi{ref})\simeq 967$~MeV
are suppressed with 
respect to the perturbative prediction. 
}

\la{fig:GStheta}
\end{figure*}

The NLO result for $G^{ }_\rmii{S}$, normalized to 
$G_{\theta}^\rmi{free}$, is shown in \fig\ref{fig:GStheta}. 
With this normalization the agreement at small $\tau$ and the discrepancy
at large $\tau$ of the NLO expression and lattice data
becomes clearly visible. Independently of lattice data, 
it is also seen that the relative magnitude of the scalar density 
correlator in the infrared domain $\tau \sim \beta/2$ 
is largest for $m(\bmu_\rmi{ref}) \approx 1.0 - 1.25$~GeV, quite
close to the experimental value
$m_c(\bmu^{ }_\rmi{ref}) = 1.275(25)$~GeV~\cite{pdg}. 
Whether the relative effect
is equally large in $\zeta$ remains to be 
seen but we nevertheless consider the pattern 
seen in \fig\ref{fig:GStheta} to be intriguing.

%
\subsection{Normalization to a reconstructed correlator}

On the lattice side it has become fashionable to normalize thermal
imaginary-time correlators to a ``reconstructed'' correlator. Even 
though we suspect that this increases 
systematic uncertainties (cf.\ below), 
we have worked out this case as well for completeness. 

Within perturbation theory, 
the massive scalar channel spectral function at zero temperature reads
(the structure is similar to the vector channel, cf.\ ref.~\cite{old})
\ba
 && \hspace*{-0.5cm}
 \rho^\rmi{vac}_\rmii{S}(\omega) 
  = 
\theta(\omega - 2 M)   
  \frac{\CA M^2 (\omega^2 - 4 M^2)^{\fr32}}{8\pi\omega}
 + \theta(\omega - 2 M) 
 \frac{4 g^2 \CA \CF M^2}{(4\pi)^3 \omega^2} 
 \biggl\{  
 \la{GS_full_vac}
 \\ & & \!\!\!
 (\omega^2 - 2 M^2) (\omega^2 - 4 M^2)  
 L_2 \biggl( \frac{\omega - \sqrt{\omega^2 - 4 M^2}}
 {\omega + \sqrt{\omega^2 - 4 M^2}} \biggr)
 + \biggl(\fr32 \omega^4  - 2 \omega^2 M^2  - 13 M^4 \biggr)
  \,\mathrm{acosh} \biggl( \frac{\omega}{2 M} \biggr)
 \nn &  & \!\!\!
 -\; \omega (\omega^2 - 4 M^2)^{\fr12}
 \biggl[
   (\omega^2 - 4 M^2) \biggl( 
  \ln \frac{\omega (\omega^2 - 4 M^2)}{M^3}
  + \fr34 \delta \biggr)
  -\fr38 (3 \omega^2 - 14 M^2) 
 \biggr]
 \biggr\} + \rmO(g^4) \;, \nonumber 
\ea
where the function $L_2$ is defined as
\be
 L_2(x) \; \equiv \; 
 4 \, \mathrm{Li}_2 (x) + 2 \, \mathrm{Li}_2(-x)
 + [2 \ln(1-x) + \ln(1+x)] \ln x
 \;. \la{L2} 
\ee
The ultraviolet asymptotics from here is
\be
 \rho^\rmi{vac}_\rmii{S}(\omega) 
 \;\; \stackrel{\omega \gg M}{\approx} \;\;
 \frac{\CA \omega^2 M^2}{8\pi}
 \biggl\{ 1   
 + \frac{6 g^2 \CF }{(4\pi)^2}
 \biggl(
  \ln\frac{M^2}{\omega^2} - \delta + \fr32 
 \biggr) 
  + \rmO(g^4) 
 \biggr\}
 \;.
 \la{full_vacS_asympt}
\ee
We note again that in the pole mass scheme, where $M$ is constant and 
$\delta = 0$, \eq\nr{full_vacS_asympt} 
turns negative at very large $\omega$ and
the NLO correction overtakes the LO term, 
implying a breakdown
of the perturbative series. If, however, we go to the $\msbar$ scheme, 
choosing $\delta$ according to \eq\nr{m_msbar} and setting
$
 M^2 \to m^2(\bmu) 
$
according to \eq\nr{m_bmu}, we obtain another representation: 
\ba
 \rho^\rmi{vac}_\rmii{S}(\omega) 
 & \stackrel{\omega \gg M}{\approx} &
 \frac{\CA\, \omega^2 m^2(\bmu_\rmi{ref})}{8\pi}
 \biggl[
   \frac{\ln(\bmu_\rmi{ref}/\Lambdamsbar)}{\ln(\bmu/\Lambdamsbar)} 
 \biggr]^{\frac{18 \CF}{11\CA - 4 \TF}}
 \nn & \times & 
 \biggl\{ 1   
 + \frac{6 g^2(\bmu) \CF }{(4\pi)^2}
 \biggl(
  \ln\frac{\bmu^2}{\omega^2} + \frac{17}{6}
 \biggr) 
 + \rmO(g^4)
 \biggr\}
 \;.
 \la{full_vacS_asympt_2}
\ea
Taking $\bmu$ to scale with $\omega$, a series is obtained which 
becomes increasingly convergent as $\omega$ grows. The result 
is independent of the precise choice of $\bmu$ up to higher-order
corrections; in practice we set\footnote{
 The infrared cutoff $\pi T$ only affects the smallest masses, 
 $m(\bmu^{ }_\rmi{ref}) \, \lsim\,  2\pi T$.
 } 
\be
 \bmu \to {\rm max}(\pi T, \omega \, e^{-17/12})
 \;.
\ee
Subsequently the ``reconstructed'' correlator is defined
by taking the vacuum spectral function 
and using it to compute a thermal Euclidean correlator from
\be
  G^\rmi{rec}_\rmii{S}(\tau) \equiv
 \int_0^\infty
 \frac{{\rm d}\omega}{\pi} \rho^\rmi{vac}_\rmii{S} (\omega)
 \frac{\cosh \left(\frac{\beta}{2} - \tau\right)\omega}
 {\sinh\frac{\beta \omega}{2}} 
 \;. \la{GS_relation}
\ee

\begin{figure*}[tb]


\centerline{%
 \epsfysize=8.5cm\epsfbox{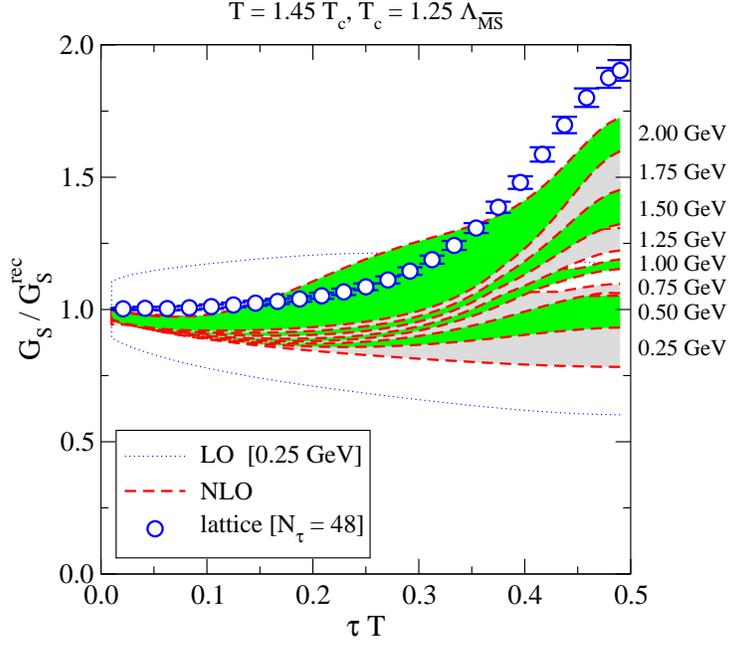}
}

\caption[a]{\small
The ratio $G_\rmii{S} / G_\rmii{S}^\rmi{rec}$ 
(cf.\ \eq\nr{GS_relation})
as a function of $\tau T$
and the quark mass $m(\bmu_\rmi{ref})$, indicated next to the coloured
bands. The bands reflect the uncertainty related to the choice of 
the renormalization scale. 
The lattice simulations are from ref.~\cite{ding2}.
The comparison contains
uncontrolled uncertainties from both sides as discussed in the text. 
}

\la{fig:GSrec}
\end{figure*}

In \fig\ref{fig:GSrec} results
normalized to the reconstructed correlator are shown. 
As expected, the results approach unity at short distances for all masses
(the small-$\tau$ result 
is determined by large $\omega$ and then thermal
corrections are strongly suppressed with respect to 
the vacuum term~\cite{simon,nlo}). However the match to lattice
data is not as good as in \figs\ref{fig:GSfree}, \ref{fig:GStheta}; 
among possible explanations we can envisage the following: 
\bi

\item
On the continuum side it is questionable whether perturbation theory, 
which misses all resonance contributions, can reflect the vacuum 
scalar spectral function even qualitatively
in the $\omega$-range dominating \eq\nr{GS_relation} at moderate $\tau$,
$\omega \,\gsim\, \pi T \sim 1$~GeV. 

\item
On the lattice side the theoretical formulae 
(e.g.\ \eq\nr{GS_relation})
used for relating 
measured data to the reconstructed 
correlator assume a continuous $\tau$-variable; this necessitates 
taking a continuum limit which has not been reached.  

\item
The ``reconstructed'' correlator of ref.~\cite{ding2} was not
measured at $T= 0$ but at $T = 0.73 \Tc$.

\ei
Because of these issues we refrain from speculating on the 
origins of the discrepancy in \fig\ref{fig:GSrec}.

%
\section{Conclusions}
\la{se:concl}

The purpose of this paper has been to investigate the influence
of a finite charm quark mass on a 2-point
imaginary-time correlator in the so-called bulk
(or scalar) channel, corresponding to the trace of the energy-momentum 
tensor in continuum QCD.  
We could have anticipated a rather substantial influence, 
given that in massless QCD the corresponding correlator is non-zero
only because of an ``anomalous'' breaking of scale invariance, and  
is therefore proportional to a high power of running coupling constants
(cf.\ \eq\nr{rho_theta_lo}).  

Remarkably, the relative influence of a massive quark on the scalar
channel correlator at large imaginary-time separations appears
to peak for  
$m(\bmu_\rmi{ref}) = 1.0 - 1.25$~GeV (cf.\ \fig\ref{fig:GStheta}),  
not far from the physical charm mass
$m_c(\bmu^{ }_\rmi{ref}) = 1.275(25)$~GeV~\cite{pdg}. 
Hopefully this observation motivates a perturbative determination 
of the bulk viscosity $\zeta$ for all quark masses 
as well as  
refined lattice investigations of scalar correlators 
in the future, including a systematic approach to the continuum limit.

An interesting principal application of the 
scalar correlator is that 
it can be used for determining the charm quark chemical 
equilibration rate close to equilibrium~\cite{chemical}, a quantity
that experiences a non-trivial (Sommerfeld) enhancement even deep
in the non-relativistic regime~\cite{sommerfeld}.  
The determination poses however 
a numerical challenge because physics
resides in a very narrow transport peak 
which is difficult to resolve from Euclidean data. 
(For $M \gg \pi T$ the peak is exponentially narrow,
cf.\ \eqs\nr{G_chem}, \nr{G_chem_2}, 
rather than only $1/M$-suppressed like in the vector channel~\cite{tp}; 
however for $ m(\bmu_\rmi{ref}) \sim 1$~GeV and 
a temperature $T \gsim 400$~MeV the difference might not 
be dramatic. Note also that there are significant challenges
in resolving the transport peak even in pure Yang-Mills
theory~\cite{ms}, however with connected mesonic correlators
a higher numerical precision can typically be reached.)

Taking the current lattice data at face value, we find discrepancies 
much larger than in the vector channel where an identical
data set was used
(cf.\ \figs\ref{fig:GSfree}, \ref{fig:GStheta} vs.\ 
ref.~\cite{GVtau}). Part of the reason
is surely that the quark mass, with associated uncertainties in its
determination, now plays a more prominent role. 
In order to get the uncertainties under control a continuum 
limit needs to be taken.  
Nevertheless it is also tempting to speculate that the scalar
channel might experience 
larger non-perturbative effects than the vector one.
Of course, these may partly originate from quarkonium physics
rather than the transport regime, cf.\ e.g.\ ref.~\cite{ding2}
and references therein.

For the full trace of \eq\nr{trace_anomaly}, 
an open challenge both on the lattice and in continuum
is the study of the ``mixed'' channel, 
with correlators involving crossterms between mesonic and purely 
gluonic operators. 
On the lattice this correlator might be rather
noisy numerically, but hopefully suitable methods will eventually 
be developed. A similar problem
concerns the disconnected contraction of mesonic scalar densities.

It may finally be wondered why we have concentrated on the bulk rather
than the more prominent shear channel. The reason is that 
for bulk viscosity the effects from massless QCD are ``anomalously''
suppressed, so that the contribution from the explicit breaking
of scale invariance through quark masses is relatively speaking more
important. In the shear channel the charm mass also has an influence, 
however we do not expect it to be larger than 
$\sim 10$\% as has been observed in basic thermodynamic 
quantities at $T \gsim 400$~MeV before~\cite{phenEOS}--\cite{buwu}.

%
\section*{Note added}

We have recently explored possible reasons for the discrepancy
between NLO expressions and lattice data in the scalar channel, by
varying the corresponding spectral function, and found that the
discrepancy is likely to originate from quarkonium-related physics,
more precisely from the proper location of the quark-antiquark
threshold~\cite{new}.

%
\section*{Acknowledgements}

We thank H.-T.~Ding for helpful discussions and 
providing us with lattice data from ref.~\cite{ding2},
and G.D.~Moore for helpful discussions.  
This work was partly supported by the Swiss National Science Foundation
(SNF) under the grant 200021-140234 as well as under 
the Ambizione grant PZ00P2-142524. 

%
\appendix
\renewcommand{\thesection}{Appendix~\Alph{section}}
\renewcommand{\thesubsection}{\Alph{section}.\arabic{subsection}}
\renewcommand{\theequation}{\Alph{section}.\arabic{equation}}

%
\section{Master sum-integrals}

The way that the computation is organized is by first using 
substitutions of sum-integration variables in order to express
the result in terms of a small number of ``master'' sum-integrals,
or basis functions, 
cf.\ \eq\nr{GS_nlo}. After transformation to coordinate space 
the basis functions can be defined as 
\be
 \mathcal{I}^{m_1m_2m_3}_{n_1n_2n_3n_4n_5} (\tau) \equiv
 T \sum_{\omega_n} e^{-i \omega_n \tau}
 \left. 
 \Tint{K\{P\}}
  \frac{(M^2)^{m_1} (Q^2)^{m_2}(2K\cdot Q)^{m_3}}
 {(K^2)_{ }^{\raise0.4ex\hbox{$\scriptstyle n_1$}} 
 \Delta^{n_2}_P \Delta^{n_3}_{P-K}\Delta^{n_4}_{P-Q}
 \Delta^{n_5}_{P-K-Q}}
 \right|_{Q = (\omega_n,\vec{0})}
 \;. \la{def_I}
\ee
In ref.~\cite{GVtau} expressions obtained after carrying out 
the Matsubara sums were given for all the basis functions
appearing in the present computation, and these have been 
employed in order to arrive at 
\eqs\nr{GS_NLO_tau}, \nr{GS_NLO_const} below. Here we just 
remark, for completeness, that one of 
the basis functions discussed in ref.~\cite{GVtau} 
is actually {\em not} independent of the others: 
\ba
 \mathcal{I}^{000}_{-11111}(\tau) & = & 
 2\, \mathcal{I}^{000}_{01110}(\tau) - 
 2\, \mathcal{I}^{100}_{01111}(\tau) - 
 \fr12\, \mathcal{I}^{010}_{01111}(\tau) 
 \;. \la{m19}  
\ea

%
\section{Renormalization of the scalar channel correlator}

Expressing \eq\nr{GS_nlo} in terms of the basis functions listed in 
appendix~A of ref.~\cite{GVtau}, we obtain results for the scalar
correlator in which the Matsubara sums have been carried out: 
\ba
 & & \hspace*{-1cm}
 \frac{ 
  \left. G_\rmii{S}^\rmii{NLO} \right|_\rmii{$\tau$-dep.}
  }{ 
  4 g^2 \CA \CF M^2
  } 
 \nn
 & = &
 \int_{\vec{p,k}} 
 \frac{D^{ }_{\epsilon_k E_p E_{pk}}(\tau) M^2}{\epsilon_k E_p E_{pk}
 \Delta^{ }_{+-}\Delta^{ }_{-+} }
 \biggl[
  \frac{\epsilon_k^2 + (E_p + E_{pk})^2 }{4 M^2}
  +  
  \frac{\Delta^{ }_{--}}{\Delta^{ }_{++}}
 - \frac{\epsilon_k^2}{\Delta^{ }_{+-}\Delta^{ }_{-+}}
 + \frac{4\epsilon_k^2 M^2}{\Delta_{++}^2 \Delta^{ }_{+-}\Delta^{ }_{-+}}
 \biggr]
 \nn 
 & + &
 \int_{\vec{p,k}} 
 \frac{D^{\epsilon_k}_{E_p E_{pk}}(\tau) M^2}{\epsilon_k E_p E_{pk}
 \Delta^{ }_{+-}\Delta^{ }_{-+}
 }
 \biggl[ 
  \frac{\epsilon_k^2 +  (E_p +  E_{pk})^2}{4 M^2}
  + 
  \frac{\Delta^{ }_{++}}{\Delta^{ }_{--}}
 - \frac{\epsilon_k^2}{\Delta^{ }_{+-}\Delta^{ }_{-+}}
 + \frac{4\epsilon_k^2 M^2}{\Delta_{--}^2 \Delta^{ }_{+-}\Delta^{ }_{-+}}
 \biggr]
 \nn 
 & - &
 \int_{\vec{p,k}} 
 \frac{2 D^{E_p}_{\epsilon_k E_{pk}}(\tau)M^2}{\epsilon_k E_p E_{pk}
  \Delta^{ }_{++}\Delta^{ }_{--}
 }
 \biggl[
   \frac{\epsilon_k^2 +  (E_p -  E_{pk})^2}{4M^2}
  + 
  \frac{\Delta^{ }_{+-}}{\Delta^{ }_{-+}}
 - \frac{\epsilon_k^2}{\Delta^{ }_{++}\Delta^{ }_{--}}
 + \frac{4\epsilon_k^2 M^2}{\Delta_{-+}^2 \Delta^{ }_{++}\Delta^{ }_{--}}
 \biggr]
 \nn 
 & + & \int_\vec{p} D^{ }_{2E_p}(\tau) \biggl\{ \mbox{``\eq\nr{coeff}''}
 \nn 
 &  &  
 \; + \, 
 \int_\vec{k} \frac{\nB{}(\epsilon_k)}{\epsilon_k} \biggl[ 
    \frac{M^2}{E_p^4}
  - \frac{M^2(E_p^2 - M^2)}{2E_p^3E^{ }_{pk}}
  \biggl( \frac{1}{\Delta_{++}^2} 
        + \frac{1}{\Delta_{--}^2}
        - \frac{1}{\Delta_{+-}^2} 
        - \frac{1}{\Delta_{-+}^2}  \biggr)
 \nn 
 &  & \qquad\qquad + \,  
 \frac{E_p^2 + E_{pk}^2 - M^2}{E_p E^{ }_{pk}}
 \biggl( 
 \frac{1}{\Delta^{}_{++}\Delta^{ }_{--}} 
 - 
 \frac{1}{\Delta^{}_{+-}\Delta^{ }_{-+}} 
 \biggr)
 \nn 
 &  & \qquad\qquad + \,  
 \frac{M^2 (2 E_p^2 - M^2)}{E_p^4}
 \biggl( 
 \frac{1}{\Delta^{}_{++}\Delta^{ }_{--}}
 + 
 \frac{1}{\Delta^{}_{+-}\Delta^{ }_{-+}}
 \biggr)
 \biggr]
 \nn &  & \; + \,  
 \int_\vec{k} \frac{\nF{}(E_{pk})}{E_{pk}} \, \mathbbm{P}\biggl[ 
  \frac{M^2}{E_p^4}
  - \frac{M^2(E_p^2 - M^2)}{2 \epsilon^{ }_k E_p^3}
  \biggl( \frac{1}{\Delta_{-+}^2} 
        + \frac{1}{\Delta_{--}^2}
        - \frac{1}{\Delta_{+-}^2}
        - \frac{1}{\Delta_{++}^2} 
  \biggr)
 \nn &   & 
 \qquad\qquad 
  + \, 
 \frac{E_p^2 + E_{pk}^2 - M^2}{ E_p} 
 \biggl(
   \frac{1}{\Delta^{ }_+\Delta^{ }_{++}\Delta^{ }_{--}}
   + \frac{1}{\Delta^{ }_-\Delta^{ }_{+-}\Delta^{ }_{-+}}
 \biggr)
 -  \frac{M^2}
         {E_p^2\, \Delta^{ }_+ \Delta^{ }_-} 
 \nn &   & \qquad\qquad + \,  
 \frac{E_{pk} M^2 (2 E_p^2  - M^2)}{E_p^4} 
 \biggl(
   \frac{1}{\Delta^{ }_+\Delta^{ }_{++}\Delta^{ }_{--}}
   - \frac{1}{\Delta^{ }_-\Delta^{ }_{+-}\Delta^{ }_{-+}}
 \biggr)
 \biggr]
 \biggr\} 
 \nn 
 & + &  \int_\vec{p} E_p \partial^{ }_{E_p} D^{ }_{2E_p}(\tau) 
 \biggl\{ 0 
 \nn & & \; + \, 
 \int_\vec{k} \frac{\nB{}(\epsilon_k)}{\epsilon_k} \biggl[ 
 \frac{1}{2 E_p^2}
 -\frac{M^2}{2 E_p^4} 
  -  
 \frac{M^2(E_p^2 - M^2)} {2 E_p^4}
 \biggl( 
   \frac{1}{ \Delta^{ }_{++}\Delta^{ }_{+-} }
 +
   \frac{1}{ \Delta^{ }_{--}\Delta^{ }_{-+}}
 \biggr)
 \biggr]
 \nn &  & \; + \,  
 \int_\vec{k} \frac{\nF{}(E_{pk})}{E_{pk}} \biggl[ 
 \frac{1}{2 E_p^2} - \frac{M^2}{2 E_p^4}
 -
 \frac{M^2(E_p^2 - M^2)}{2E_p^4}
 \biggl( 
   \frac{1}{ \Delta^{ }_{++}\Delta^{ }_{--} }
 +
   \frac{1}{ \Delta^{ }_{+-}\Delta^{ }_{-+}}
 \biggr)
 \biggr]
 \biggr\} 
 \;. \la{GS_NLO_tau}
\ea
The constant contribution, in turn, can be expressed as 
\ba
 & & \hspace*{-1cm}
 \frac{ 
  \left. G_\rmii{S}^\rmii{NLO} \right|_\rmii{const.}
  }{ 
  4 g^2 \CA \CF M^2
  } 
 \nonumber \\[2mm]
 & = &
 \int_{\vec{p,k}} 
 \frac{T \nF{}'(E_p)\nF{}'(E_{pk})}{2 E_p^2 E_{pk}^2}
 \biggl\{
     - 2M^2
     + 2M^4 \biggl( 
   \frac{1}{ \Delta^{ }_{++}\Delta^{ }_{--} }
 +
   \frac{1}{ \Delta^{ }_{+-}\Delta^{ }_{-+}}
 \biggr)
 \biggr\}
 \nn 
 & + & \int_\vec{p} 2 T \nF{}'(E_p) \biggl\{ \mbox{``\eq\nr{coeff2}''}
 \nn & & \; + \, 
 \int_\vec{k} \frac{\nB{}(\epsilon_k)}{\epsilon_k} \biggl[ 
  \frac{M^2}{E_p^4}
 - 
 \frac{2 \epsilon_k M^4}{E_p^3 }
 \biggl(
     \frac{1}{\Delta_{++}^2\Delta_{+-}^2}
  - 
     \frac{1}{\Delta_{--}^2\Delta_{-+}^2}
 \biggr)
 \nn 
 &  & \qquad\qquad + \, 
 \frac{M^2(2E_p^2 - M^2)}{E_p^4 }
 \biggl(
    \frac{1}{\Delta^{ }_{++}\Delta^{ }_{+-}} 
  +
    \frac{1}{\Delta^{ }_{--}\Delta^{ }_{-+}} 
 \biggr)
 \biggr]
 \nn &  & \; + \, 
 \int_\vec{k} \frac{\nF{}(E_{pk})}{E_{pk}} \biggl[ 
   \frac{M^2}{E_p^4}
  + \frac{M^2}{E_p^2 E_{pk}^2}
 + 
  \frac{2M^4}{ E_p^3 E^{ }_{pk} }
 \biggl(
    \frac{\Delta_+^2}{\Delta_{++}^2\Delta_{--}^2}
  -  \frac{\Delta_-^2}{\Delta_{+-}^2\Delta_{-+}^2}
 \biggr)
 \nn 
 &  & \qquad\qquad + \, 
 \biggl( 
 \frac{M^2 ( 2 E_p^2 -M^2)} {E_p^4} 
  - \frac{M^4} {E_p^2 E_{pk}^2} 
 \biggr)
 \biggl( \frac{1}{\Delta^{ }_{++} \Delta^{ }_{--}}
 + \frac{1}{\Delta^{ }_{+-} \Delta^{ }_{-+}} \biggr)
 \biggr]
 \biggr\} 
 \nn 
 & + & \int_\vec{p} 2 T E_p \nF{}''(E_p) \biggl\{ 0 
 \nn & & \; + \, 
 \int_\vec{k} \frac{\nB{}(\epsilon_k)}{\epsilon_k} \biggl[ 
 -\frac{M^2}{2 E_p^4} 
  +      
 \frac{M^4}{2 E_p^4} 
 \biggl( 
   \frac{1}{ \Delta^{ }_{++}\Delta^{ }_{+-} }
 +
   \frac{1}{ \Delta^{ }_{--}\Delta^{ }_{-+}}
 \biggr)
 \biggr]
 \nn &  & \; + \,  
 \int_\vec{k} \frac{\nF{}(E_{pk})}{E_{pk}} \biggl[ 
 -\frac{M^2}{2 E_p^4} 
  +      
 \frac{M^4}{2 E_p^4} 
 \biggl( 
   \frac{1}{ \Delta^{ }_{++}\Delta^{ }_{--} }
 +
   \frac{1}{ \Delta^{ }_{+-}\Delta^{ }_{-+}}
 \biggr)
 \biggr]
 \biggr\} 
 \;. \la{GS_NLO_const}
\ea

The ``0''s in \eqs\nr{GS_NLO_tau}, \nr{GS_NLO_const}
represent vacuum contributions that vanish after renormalization. 
The coefficients of $D_{2E_p}(\tau)$ and $T\nF{}'(E_p)$  are also 
related to renormalization, but do not vanish:  
\ba
 \mbox{``\eq\nr{coeff}''} & = & \int_\vec{k} \mathbbm{P}\biggl\{ 
  \frac{ (1-\epsilon)(2M^2-E_p^2) }{2 E_p^2 M^2}
  \biggl(\frac{1}{\epsilon^{ }_k} - \frac{1}{E^{ }_{pk}} \biggr)
 - 
  \frac{
     \epsilon E_p^2 + (1-\epsilon) M^2    
  }{2 E_p^2 E^{ }_{pk}
        (E_{pk}^2 - E_p^2)}
 \nn & & +\, 
 \frac{\epsilon\, \epsilon_k + E_{pk}} 
    {\epsilon_k E_{pk}[(\epsilon_k + E_{pk})^2 - E_p^2]}
 + 
  \frac{2 M^2  (\epsilon_k + E_{pk}) (E_p^2 - M^2 )}
         {\epsilon_k E^{ }_{pk} E_p^2
         [(\epsilon_k + E_{pk})^2 - E_p^2]^2} 
 \nn & & 
   + \,\frac{
       (\epsilon_k + 2 E^{ }_{pk})
       (E_p^2 - M^2)
       (2E_p^2 - M^2)
        }
        {\epsilon_k E^{ }_{pk} E_p^2
         [(\epsilon_k + E_{pk})^2 - E_p^2](E_{pk}^2 - E_p^2)}
 \biggr\} \;, \hspace*{1cm} \la{coeff} \\
 \mbox{``\eq\nr{coeff2}''} & = & 
  \int_\vec{k} \mathbbm{P} \biggl\{ 
   \frac{1-\epsilon}{E_p^2}
   \biggl( \frac{1}{\epsilon^{ }_k } -  \frac{1}{ E^{ }_{pk} }  \biggr)
   -\frac{(1 - \epsilon)M^2} { 2 E_p^2 E^3_{pk} } 
   \nn & & 
   +\,   \frac{[-E_p^2 +  (\epsilon_k + E_{pk}) (\epsilon_k + 3 E_{pk}) ] M^4}
         { E_p^2 E_{pk}^3 
         [(\epsilon_k + E_{pk})^2 - E_p^2]^2} 
 \biggr\} \;. \hspace*{1cm} \la{coeff2} 
\ea
Reducing to a basis of independent structures as indicated
in \eqs(B.17)--(B.20) of ref.~\cite{GVtau}, and carrying out
the integrals in the structures with one or two propagators, we obtain
\ba
 & & \hspace*{-9mm} \mbox{``\eq\nr{coeff}''}  =  
 \nonumber \\[4mm] 
 & & \hspace*{-9mm}
 \biggl\{\int_K \mathbbm{P} \, \biggl[ 
   2 \biggl( 1 - \frac{M^2}{E_p^2} \biggr)
   \biggl( \frac{1-\epsilon}{M^2\Delta^{ }_K} - \frac{1-\epsilon}{M^2K^2}
   + \frac{1}{\Delta_K^2} 
     - \frac{\epsilon}{\Delta^{ }_K \Delta^{ }_{K-Q}}
   \biggr)
 \nn & & 
   +\, \frac{2 M^2}{E_p^2} \biggl(  
   \frac{1}{K^2\Delta^{ }_{P-K}} 
     - \frac{1}{\Delta^{ }_K \Delta^{ }_{K-Q}}
   \biggr)
 \nn & & 
   + \, \biggl(1 - \frac{M^2}{E_p^2}\biggr) 
      \frac{4 M^2}{K^2 \Delta^2_{P-K}} 
   +  \biggl( 1 - \frac{M^2}{E_p^2} \biggr)
    \frac{4(2 E_p^2 - M^2)}{K^2\Delta^{ }_{P-K}\Delta^{ }_{P-K-Q}}
 \biggr]\biggr\}_{p_0 = i E_p,Q = (2 i E_p,\vec{0})} + \rmO(\epsilon)
 \nn & = & \frac{1}{8\pi^2} \biggl[ 
 \frac{M^2}{E_p^2}
 \biggl( 1 + \frac{p}{E_p} \ln\frac{E_p + p}{E_p - p}\biggr)
 - 1 
 \biggr]  
 \nn & & \; + \,   
 \biggl(1  - \frac{M^2}{E_p^2} \biggr) \int_\vec{k} \, \mathbbm{P}
 \biggl\{ 
   - \frac{M^2}{ E_{pk} }
    \biggl[
      \frac{1}{2 \epsilon^{ }_k (\epsilon^{ }_k + E^{ }_p) \Delta_{++}^2}
     + \frac{1}{2 \epsilon^{ }_k (\epsilon^{ }_k - E^{ }_p) \Delta_{-+}^2}
     - \frac{1}{(\epsilon_k^2 - E_p^2 )E_{pk}^2} 
    \biggr]
 \nn & & \qquad\qquad\qquad - \, 
 \frac{2 E_p^2 - M^2}
 {2\epsilon_k^2 E^{ }_p E^{ }_{pk}}
    \biggl(
      \frac{1}{\Delta_{+}^{ }} 
     + \frac{1}{\Delta_{-}^{ }} 
     -  \frac{1}{\Delta_{++}^{ }} 
     +  \frac{1}{\Delta_{-+}^{ }} 
    \biggr)
 \biggr\} + \rmO(\epsilon)
 \;,  \la{GS_div_tau} \\
 & & \hspace*{-9mm} \mbox{``\eq\nr{coeff2}''}  =  
 \nonumber \\[4mm] 
 & & \hspace*{-9mm}
 \biggl\{\int_K \mathbbm{P} \, \biggl[ 
   \frac{2(1-\epsilon)}{E_p^2}
  \biggl( \frac{1}{K^2} - \frac{1}{\Delta^{ }_K}\biggr)
   +\frac{2M^2}{E_p^2 K^2\Delta^{ }_{P-K}} 
   - \frac{2 (2- \epsilon)M^2}{E_p^2 \Delta_K^2}
 \biggr]
 \biggr\}_{p_0 = i E_p,Q = (2 i E_p,\vec{0})} + \rmO(\epsilon)
 \nn & = & \frac{3}{8\pi^2} \frac{M^2}{E_p^2}
 + \rmO(\epsilon)
 \;. \la{f_coeff2}
\ea
In the $\vec{k}$-integral of \eq\nr{GS_div_tau} 
the angular integration can be carried out, which leads to the vacuum 
part of \eq\nr{GS_NLO_tau_final}; 
\eq\nr{f_coeff2} in turn 
yields the vacuum part of \eq\nr{GS_NLO_const_final}. 
Note that both \eq\nr{GS_div_tau} and \nr{f_coeff2} contain
a counterterm contribution from \eq\nr{GS_ct} and are 
therefore finite in the ultraviolet regime (large $k$). 

%

\end{document}